# Study on Downlink Spectral Efficiency in Orthogonal Frequency Division Multiple Access Systems

Qiang Huo, Meng Ma, and Bingli Jiao

**Abstract**

In previous studies on the capacity of orthogonal frequency division multiple access (OFDMA) systems, it is usually assumed that co-channel interference (CCI) from adjacent cells is a Gaussian-distributed random variable. However, very-little work shows that the Gaussian assumption does not hold true in OFDMA systems. In this paper, the statistical property of CCI in downlink OFDMA systems is studied, and spectral efficiency of downlink OFDMA system is analyzed based on the derived statistical model. First, the probability density function (PDF) of CCI in downlink OFDMA cellular systems is studied with the considerations of path loss, multipath fading and Gaussian-like transmit signals. Moreover, some closed-form expressions of the PDF are obtained for special cases. The derived results show that the PDFs of CCI are with a heavy tail, and significantly deviate from the Gaussian distribution. Then, based on the derived statistical properties of CCI, the downlink spectral efficiency is derived. Numerical and simulation results justify the derived statistical CCI model and spectral efficiency.

The authors are with the School of Electronics Engineering and Computer Science, Peking University, Beijing 100871, China (e-mail: {qiang.huo, mam, jiaobl}@pku.edu.cn).

The work was partially supported by the National Nature Science Foundation of China under grant number 61101080, and Beijing Higher Education Young Elite Teacher Project.

Digital Object Identifier 10.1049/iet-com.2013.0972

# I. INTRODUCTION

Owing to the scarceness of wireless spectrum, frequency reuse is one of the fundamental approaches to achieve high capacity in cellular systems [1]. Recently, utilizing frequency reuse scheme with a smaller frequency reuse factor has been known as an attractive technique in orthogonal frequency division multiple access (OFDMA)-based fourth generation (4G) systems because of its high spectral efficiency [2]. In this case, the performance in downlink transmission, especially when the mobile station (MS) receiver is near to the edge of a cell, is mainly limited by the amount of co-channel interference (CCI) [3].

In previous studies, CCI is usually modeled as a Gaussian variable. In [4]–[7], it is assumed that CCI from adjacent cells is a Gaussian-distributed random variable. The theoretical foundation of the Gaussian assumption is the Central Limit Theorem, which can be satisfied for a large number of independent interferences, however, may does not hold true in practice. In [8]–[12], based on a condition that the receiver has full instantaneous channel-state information (CSI) of all the CCI channels, the distribution of CCI is also Gaussian-like, because that the transmit signal is Gaussian-distributed and the channel fading coefficient is a constant for the receiver. However, to estimate all the CCI channels will lead to extra pilot overhead and complexity at the receiver.

Very-little work shows that downlink CCI is non-Gaussian [13], [14]. In [13], the authors study the CCI powers for a downlink cellular communication with full loading by using a simulation approach, and simulation results show that the Gaussian assumption of CCI is a poor approximation in the realistic environments and thus a more precise non-Gaussian distributed interference model is needed. In [14], the authors study the statistical model of inter-cell interference for downlink OFDMA cellular networks, and simulation results show that as the cell loading decreases, the distribution of the interference signal deviates significantly from the Gaussian distribution.

The studies on spectral efficiency for orthogonal frequency division multiplexing (OFDM) system are mostly based on the Gaussian interference model so far. In [4]–[7], the authors analyze the spectral efficiency with the Gaussian approximation of CCI. In [8]–[12], the studies on the spectral efficiency are with the assumption of full instantaneous CSIs of all the CCI channels at the receiver.

In this paper, we study the spectral efficiency of downlink cellular systems. First, the probability density function (PDF) of CCI is derived using the characteristic function (CF) with the considerations of path loss, multipath fading and Gaussian-like transmit signals. The derived PDF is observed to highly deviate from the Gaussian distribution and possesses a heavier tail. Based on the derived statistical model of CCI, the spectral efficiency is analyzed. Simulation results justify our derived statistical CCI model and the spectral efficiency evaluation.

The rest of the paper is organized as follows. In Section II, The system model is described. The statistical model of CCI in downlink OFDMA cellular systems is presented in Section III. In Section IV, the spectral efficiency is analyzed based on the derived statistical model. Simulation results are presented in Section V. Section VI concludes the paper.

*Notations*: $\Pr[\cdot]$ and $\mathrm{E}[\cdot]$ denote probability and expectation, respectively. The universal quantifier symbol $\forall$ is used to indicate universal quantification.

## II. System Model

In this paper, we consider a synchronous, downlink, hexagonal OFDMA cellular system consisting of $M+1$ base stations (BSs) and one MS. An example of cellular system with $M=6$ is illustrated in Fig. 1. We should emphasize that our analysis in this paper holds for arbitrary $M$, and let $BS_0$ denote the desired BS and the other $M$ BSs are interfering BSs. It is assumed that the received signal at each MS in a cell is interfered by the active BSs in other cells.

As the real and imaginary part of the received signal are two independent and identically distributed random variables (i.i.d. RVs), for the sake of simplicity, we only consider the real part of the signal in the following analysis. Thus we can assume that the BSs transmit real-valued Gaussian-like signals in the available subcarriers in each OFDM symbol [4], [11], [12] and the fading coefficients are also real-valued and Gaussian-like. As commonly assumed, the length of the cyclic prefix (CP) of an OFDM symbol is larger than the maximum multi-path delay, and thus no inter-symbol interference (ISI) occurs during the demodulation of the OFDM symbols [15], [16]. Also, it is assumed that the Doppler spread is negligible, which is usually valid in the cases of relatively slowly moving MSs, and thus the channel remains constant during one frame [1]. Finally, the desired MS is assumed to be perfectly synchronized with the target BS, and the frequency reuse factor is equal to one in this paper. In the following section, the statistical model of CCI is studied, and then the spectral efficiency is analyzed using the derived statistical CCI model.

As in our analysis, the signals and channel parameters on different subcarriers and time slots follow the same statistical properties, the indices of both subcarrier and time slot are ignored. Let $Y$ stand for the received signal at the desired MS in cell 0. The received signal of the desired MS on a specific subcarrier is given by

$$Y = S + \sum_{m=1}^{M} I_m + N = S + Z \,, \tag{1}$$

where $S = \sqrt{E_0} H_0 X_0$ denotes the received desired signal from $BS_0$, $I_m = \Phi_m \sqrt{E_m} H_m X_m$ denotes the CCI from the $m$th interfering BS, $N$ denotes the additive white Gaussian noise (AWGN) with zero mean and variance $\sigma_N^2$. $Z = \sum_{m=1}^{M} I_m + N = \sum_{m=1}^{M} \Phi_m \sqrt{E_m} H_m X_m + N$ represents the effect of CCI and AWGN, where $E_m$ represents the average received signal power at the MS from the $m$th BS, $H_m$ denotes the fading coefficient between the $m$th BS and the desired MS, and $X_m$ represents the transmit symbol from the $m$th BS, respectively. It is assumed that $H_m$ is stationary, ergodic random process with $\mathrm{E}\left[H_m^2\right]=1$. Let $X_m$, $m=0,1,\cdots,M$ are i.i.d. RVs with $\mathrm{E}\left[X_m^2\right]=1$. $\Phi_m$ is equal to 1 if the corresponding subcarrier is occupied by the *m*th BS and 0 otherwise. In this paper, we assume that the subcarriers are assigned to the users with equal probability and equal transmit power. The loading rate of the subcarriers is denoted by $p_m$ for the *m*th BS, i.e., $\Pr\left[\Phi_m=1\right]=p_m$ and $\Pr\left[\Phi_m=0\right]=1-p_m$, and $P_m$ represents the transmit power of the *m*th BS, and $\gamma_m=d_m^{-\alpha_m}$ denotes the effect of the path loss between the *m*th BS and the desired MS, where $d_m$ is the distance between the *m*th BS and the MS, and $\alpha_m$ denote the corresponding path loss exponent. Thus the average received signal power, $E_m$, can be represented as $E_m=P_m\gamma_m$.

## III. STATISTICAL MODELS OF CCI AND RECEIVED SIGNAL

In this section, the statistical models of CCI and received signal are derived. Firstly, in Subsection III-A, generalized expressions of the PDFs of received CCI and total received signal are derived by using the CF. Secondly, closed-form expressions of the CCI's PDFs are presented for some special cases in Subsection III-B. Finally, some numerical and analytical results of the distributions of CCI are presented in Subsection III-C.

### *A. Generalized PDFs of CCI and Total Received Signal*

The PDF and the corresponding CF of the CCI from *m*th BS, i.e., $I_m$, are derived in Appendix A as

$$f_{I_m}(x)=\frac{p_m}{\pi\sqrt{E_m}}K_0\left(\frac{|x|}{\sqrt{E_m}}\right)+(1-p_m)\delta(x), \tag{2}$$

and

$$\Psi_{I_m}(w)=p_m\left(\frac{1}{1+E_m w^2}\right)^{1/2}+(1-p_m), \tag{3}$$

respectively, where $m=1,\cdots,M$ is the BS index, $\mathrm{K}_0(x)$ is the modified Bessel function of the second kind [17, Sec 10.25], and $\delta(x)$ is the Dirac delta function [18, Sec 19.1.3].

Since the transmit signal and channel coefficient of each BS are independent with each other, the CF of the PDF of the total CCIs, denoted by $I=\sum_{m=1}^{M} I_m$, is equal to the product of the CF of each independent interference [19, Sec 7.2], and can be expressed as

$$\Psi_I(w)=\prod_{m=1}^{M}\Psi_{\mathrm{I}_m}(w)=\prod_{m=1}^{M}\left(p_m\left(\frac{1}{1+E_m w^2}\right)^{1/2}+\left(1-p_m\right)\right). \tag{4}$$

Therefore, the PDF of the total interference, $I$, can be expressed using the inverse CF transform [19, Sec 7.2, Eq. (5.66)] as

$$\begin{aligned} f_I(x) &= \frac{1}{2\pi}\int_{-\infty}^{\infty}\Psi_I(w)e^{-jwx}\mathrm{d}w \\ &= \frac{1}{2\pi}\int_{-\infty}^{\infty}\prod_{m=1}^{M}\left(p_m\left(\frac{1}{1+E_m w^2}\right)^{1/2}+\left(1-p_m\right)\right)e^{-jwx}\mathrm{d}w \end{aligned}. \tag{5}$$

As the thermal noise is a Gaussian RV with zero mean and variance $\mathrm{E}\left[N^2\right]=\sigma_N^2$, and the corresponding CF of the Gaussian noise can be expressed as $\Psi_N(w)=\exp\left(-\frac{\sigma_N^2 w^2}{2}\right)$ [19, Sec 5.5, Eq. (5.65)]. Thus, the CF of the sum of CCIs and AWGN is expressed as

$$\Psi_Z(w)=\Psi_N(w)\Psi_I(w)=\exp\left(-\frac{\sigma_N^2 w^2}{2}\right)\prod_{m=1}^{M}\left(p_m\left(\frac{1}{1+E_m w^2}\right)^{1/2}+\left(1-p_m\right)\right), \tag{6}$$

and the corresponding PDF is given by

$$\begin{aligned} f_Z(x) &= \frac{1}{2\pi}\int_{-\infty}^{\infty}\Psi_Z(w)e^{-jwx}\mathrm{d}w \\ &= \frac{1}{2\pi}\int_{-\infty}^{\infty}\exp\left(-\frac{\sigma_N^2 w^2}{2}\right)\prod_{m=1}^{M}\left(p_m\left(\frac{1}{1+E_m w^2}\right)^{1/2}+\left(1-p_m\right)\right)e^{-jwx}\mathrm{d}w \end{aligned}. \tag{7}$$

Noting that for a given $E_0$ and $H_0$, $S+N$ is a zero-mean Gaussian RV with variance $E_0\left|H_0\right|^2+\sigma_N^2$, we can derive the CF and PDF of the total received signal $Y=S+I+N$ by

simply replacing $\sigma_N^2$ with $E_0|H_0|^2+\sigma_N^2$ in the expressions of $Z=I+N$, i.e., in Eq. (6) and Eq. (7). The CF and PDF of the received signals, $Y$, conditioned on $E_0$ and $H_0$ are derived as

$$\Psi_{Y|H_0}\left(w|H_0\right)=\exp\left(-\frac{\left(E_0|H_0|^2+\sigma_N^2\right)w^2}{2}\right)\prod_{m=1}^{M}\left(p_m\left(\frac{1}{1+E_m w^2}\right)^{1/2}+\left(1-p_m\right)\right), \quad (8)$$

and

$$\begin{aligned} f_{Y|H_0}\left(x|H_0\right)&=\frac{1}{2\pi}\int_{-\infty}^{\infty}\Psi_{Y|H_0}\left(w|H_0\right)e^{-jwx}\mathrm{d}w \\ &=\frac{1}{2\pi}\int_{-\infty}^{\infty}\exp\left(-\frac{\left(E_0|H_0|^2+\sigma_N^2\right)w^2}{2}\right)\prod_{m=1}^{M}\left(p_m\left(\frac{1}{1+E_m w^2}\right)^{1/2}+\left(1-p_m\right)\right)e^{-jwx}\mathrm{d}w \end{aligned}, \quad (9)$$

respectively.

According to the derived PDF expressions in Eq. (5), Eq. (7), and Eq. (9), one can calculate the derived PDFs by using computational software.

*B. Closed-form expressions of CCI's PDFs*

In this subsection, some closed-form expressions of the CCI's PDFs are derived for some special cases. For analysis simplicity, we only consider the case of full loading rate, i.e., $p_m=1$, for $\forall m$ in this subsection.

First, we consider a simple case of only one interfering BS, i.e., $M$ = 1. The practical scenario of this case is that when the MS is at the edge of two cells, the received interference from the adjacent BS is much stronger than the interferences from other BSs, thus the other interferences can be ignored. Another potential scenario is that there is only one active interfereing BS adjacent to the MS.

Substituting $p_m=1$ into Eq. (2), the PDF of the single CCI signal is given as

$$f_1\left(x\right)=\frac{1}{\pi\sqrt{E_m}}\mathrm{K}_0\left(\frac{|x|}{\sqrt{E_m}}\right). \quad (10)$$

Note that $\mathrm{K}_0(x) \sim \ln(x)$ when $x \to 0$ and thus $\lim_{x \to 0} f_I(x) = \infty$ [17, Sec. 10.3.3]. It indicates that the distribution of single CCI signal is more centralized than the Gaussian distribution around the mean value.

In the case of two CCI signals with the same average received power, i.e., $E_1 = E_2$, the PDF of $I = I_1 + I_2$ is derived in Appendix B as

$$f_2(x) = \frac{1}{2\sqrt{E_1}} \exp\left( -\frac{|x|}{\sqrt{E_1}} \right). \tag{11}$$

Similarly, for the case of four interference signals with the same average received power, i.e., $E_1 = E_m$ and $m = 2,3,4$, the PDF of $I = \sum_{m=1}^{4} I_m$ is given in Appendix B as

$$f_3(x) = \exp\left( -\frac{|x|}{\sqrt{E_1}} \right)\left( \frac{|x|}{4E_1} + \frac{1}{4\sqrt{E_1}} \right). \tag{12}$$

In the case of six interference signals with the same average received power, i.e., $E_1 = E_m$ and $m = 2,\cdots,6$, then the PDF of $I = \sum_{m=1}^{6} I_m$ is given in Appendix B as

$$f_4(x) = \exp\left( -\frac{|x|}{\sqrt{E_1}} \right)\left( \frac{|x|^2}{16E_1\sqrt{E_1}} + \frac{3|x|}{16E_1} + \frac{3}{16\sqrt{E_1}} \right). \tag{13}$$

Thus, we derived the closed-form expressions for the cases when the MS is interfered by the adjacent $M$ = 1, 2, 4 or 6 active BSs with equal power as shown in Eq. (10), (11), (12) and (13), respectively.

Then, we consider the case when the MS moves toward the edge of three cells as shown in Fig. 1 (see the dashed line). Considering the first tier of interfering BSs, we have $E_m = E_{7-m}$ for $m = 1,2,3$. The PDF of $I = \sum_{m=1}^{6} I_m$ is derived in Appendix B as

$$f_I(x)=\sum_{m=1}^{3}\frac{a_m}{2\sqrt{E_m}}\exp\left(-\frac{|x|}{\sqrt{E_m}}\right), \quad (14)$$

where $a_1=\frac{E_1^2}{(E_1-E_2)(E_1-E_3)}$, $a_2=\frac{E_2^2}{(E_2-E_1)(E_2-E_3)}$ and $a_3=\frac{E_3^2}{(E_3-E_1)(E_3-E_2)}$.

Eq. (11)–(13) can be seen as three special cases of Eq. (14). For example, Eq. (14) reduces to Eq. (11) when $E_2=E_3=0$, and reduces to Eq. (12) when $E_3=0$ and $E_2\to E_1$, and reduces to Eq. (13) when $E_2\to E_1$ and $E_3\to E_1$.

*C. Numerical and Analytical Results*

As we mainly focus on the distribution, instead of the power of interference, in the following numerical, analytical and simulated results, all the variables are normalized to unity variance.

We consider a downlink OFDMA cellular network with the first tier of interfering BSs as shown in Fig. 1. We assume that the corresponding path loss exponent is equal to 4, i.e., $\alpha_m=\alpha=4$, $\forall m$, and the transmit power of the BSs is assumed to be equal, i.e., $P_m=P$, $\forall m$, and the loading rates of each cell are identical, i.e., $p_m=p$, $\forall m$. The PDF can be obtained by calculating Eq. (5) or the derived closed-form expression, i.e., Eq. (14), for some special cases. Another method to obtain the CCI's PDF is the Monte Carlo simulation. In each Monte Carlo simulation, we generate the interference signals and channel coefficients randomly, and sum up the *M* independent interference signals to obtain the received interference signal. After many times of simulations, we calculate the distribution of the received interference signals, and thus can derive the CCI's PDF. If the MS is at point A, as illustrated in Fig. 1, and the loading rate is $p=0.5$, we can obtain the numerical results of the CCI's PDF by calculating Eq. (5) numerically. The numerical and simulation results are shown in Fig. 2. One can find that the numerical result of Eq. (5) matches well with the simulation results. If the MS is at point B, i.e., at the edge of three cells, and the loading rate is $p=1$, the CCI's PDF is derived as Eq. (14). The analytical and simulated results are shown in Fig. 3. It is shown that the theoretical results obtained by Eq. (14) also match well with the simulation results. For comparison, the Gaussian PDF is also plotted in both Fig. 2 and Fig. 3, and one can find that the distributions of CCI deviate significantly from the Gaussian distribution and possess a heavier tail.

Next, we consider the case with $M = 1$, 2, 4 and 6 CCIs and each CCI has the same power. From Fig. 4, it can be observed that the analytical theoretical curves obtained by Eq. (10)–(13) match well with the simulated results. Also, from Fig. 4, one can find that while the number of CCIs decreases, the distribution of total CCI deviates seriously from the Gaussian distribution. Even in the case of six CCIs, the distribution of total CCI is still found to significantly deviate from the Gaussian distribution. This is because that in a cellular system, the number of adjacent active BSs is small, and the sum of a small number of independent interference does not converge to the Gaussian distribution. The results in Fig. 4 also reveal that the spectral efficiency in downlink OFDM cellular systems would deviate from the Gaussian channel capacity too.

In order to quantitatively analyze the difference between the distribution of CCI signals and the Gaussian distribution, the well-known mean, i.e., Kullback-Leibler distance, can be applied, which is defined as [20]

$$D\left( f_Z(x) \| g(x) \right) = \int f_Z(x) \log \frac{f_Z(x)}{g(x)} \mathrm{d}x , \tag{15}$$

where $f_Z(x)$ and $g(x)$ denote the PDF of CCI and Gaussian RV, respectively. The Kullback-Leibler distance of the PDF of CCI and Gaussian RV is plotted in Fig. 5 with different loading rates and MS positions. The MS moves along the solid line in Fig. 1. It can be seen that as the loading rate decreases, the distribution of CCI signal deviates significantly from the Gaussian signal. For a given loading rate, the Kullback-Leibler distance increase as the distance between the MS and the desired BS increases, which means that the deviation from Gaussian distribution increases when the MS moves towards the cell edge.

## IV. Spectral Efficiency Analysis

In this section, the maximum achievable rates of reliable communication with Gaussian-like transmit signals are presented based on the above-derived statistical models of CCI signals and total received signals in Section III.-A. For notational clarity, we denote the channel occupancy vector by $\mathbf{\Phi} = [\Phi_1, \Phi_2, \cdots, \Phi_M]$, the average received power vector by $\mathbf{E} = [E_1, E_2, \cdots, E_M]$ and the fading coefficient vector by $\mathbf{H} = [H_1, H_2, \cdots, H_M]$, respectively.

If the MS receiver has full CSI of all the desired transmission channels and the interference channels, i.e., $E_0$, $H_0$, $\mathbf{\Phi}$, $\mathbf{E}$ and $\mathbf{H}$ are perfectly known at the receiver, the spectral efficiency is [11], [12]

$$I_{\text{CSI}} = I\left(X_0; Y, E_0, H_0, \mathbf{\Phi}, \mathbf{E}, \mathbf{H}\right) = \mathrm{E}_{H_0, \mathbf{\Phi}, \mathbf{H}}\left(\frac{1}{2}\log\left(1 + \frac{E_0 |H_0|^2}{\sum_m \Phi_m E_m |H_m|^2 + \sigma_N^2}\right)\right). \tag{16}$$

If the MS receiver has no CSI of the interference channels, but only the CSI of the desired transmission channel, a conventional spectral efficiency estimation method based on the Gaussian interference assumption is [5]-[7]

$$I_{\text{GA}} = I_{\text{GA}}\left(X_0; Y, E_0, H_0\right) = \mathrm{E}_{H_0}\left(\frac{1}{2}\log\left(1 + \frac{E_0 |H_0|^2}{\sum_m p_m E_m + \sigma_N^2}\right)\right). \tag{17}$$

However, as mentioned above, in this paper, we treat the interference as a non-Gaussian variable, therefore, the spectral efficiency should be calculated by using mutual information as

$$\begin{aligned} I_{\text{P}} &= I\left(X_0; Y, E_0, H_0\right) \\ &= I\left(X_0; E_0\right) + I\left(X_0; H_0 | E_0\right) + I\left(X_0; Y | E_0, H_0\right) \\ &= 0 + 0 + I\left(X_0; Y | E_0, H_0\right) \\ &= H\left(Y | E_0, H_0\right) - H\left(Y | X_0, E_0, H_0\right) \\ &= \mathrm{E}_{Y, H_0}\left(\log \frac{1}{f_{Y|H_0}\left(Y | H_0\right)}\right) - \mathrm{E}_{Y, H_0}\left(\log \frac{1}{f_{Y|X_0, H_0}\left(Y | X_0, H_0\right)}\right) \end{aligned}. \tag{18}$$

According to Eq. (1), we have $\mathrm{E}_{Y,H_0}\left(\log\frac{1}{f_{Y|X_0,H_0}\left(Y|X_0,H_0\right)}\right)=\mathrm{E}_Z\left(\log\frac{1}{f_Z(Z)}\right)$. Thus, Eq. (18) can be rewritten as

$$
\begin{aligned}
I_{\mathrm{P}} &= \mathrm{E}_{Y,H_0}\left(\log\frac{1}{f_{Y|H_0}\left(Y|H_0\right)}\right)-\mathrm{E}_Z\left(\log\frac{1}{f_Z(Z)}\right) \\
&= \iint_{y,h_0} f_{Y,H_0}\left(y,h_0\right)\log\frac{1}{f_{Y|H_0}\left(y|h_0\right)}dydh_0 \ -\int_z f_Z(z)\log\frac{1}{f_Z(z)}dz \\
&= \int_{h_0} f_{H_0}\left(h_0\right)\int_y f_{Y|H_0}\left(y|h_0\right)\log\frac{1}{f_{Y|H_0}\left(y|h_0\right)}dydh_0 \ -\int_z f_Z(z)\log\frac{1}{f_Z(z)}dz
\end{aligned} \quad , \tag{19}
$$

where $f_{H_0}\left(h_0\right)$ is the PDF of fading coefficient of the desired signal, $f_{Y|H_0}\left(y|h_0\right)$ and $f_Z(z)$ are derived in Eq. (7) and Eq. (9), respectively.

The relationship between $I_{\mathrm{CSI}}$ and $I_{\mathrm{GA}}$ can be found by using the Jensen's inequality [20]. With the consideration of that $\log\left(1+\frac{c}{x}\right)$ is strictly concave with respect to $x$, when $c$ is a positive constant, and according to Jensen's inequality, we have

$$
\mathrm{E}_{H_0,\boldsymbol{\Phi},\mathbf{H}}\left(\frac{1}{2}\log\left(1+\frac{E_0\left|H_0\right|^2}{\sum_m \Phi_m E_m\left|H_m\right|^2+\sigma_N^2}\right)\right)>\mathrm{E}_{H_0}\left(\frac{1}{2}\log\left(1+\frac{E_0\left|H_0\right|^2}{\mathrm{E}_{\boldsymbol{\Phi},\mathbf{H}}\left(\sum_m \Phi_m E_m\left|H_m\right|^2+\sigma_N^2\right)}\right)\right), \tag{20}
$$

and, which means that $I_{\mathrm{CSI}}>I_{\mathrm{GA}}$.

In order to compare the three spectral efficiencies, we introduce the difference factor of the spectral efficiencies, $D_{\mathrm{CSI}}^{\mathrm{P}}$ and $D_{\mathrm{GA}}^{\mathrm{P}}$, which are defined as

$$
D_{\mathrm{CSI}}^{\mathrm{P}}=\frac{\left|I_{\mathrm{P}}-I_{\mathrm{CSI}}\right|}{I_{\mathrm{P}}}\times 100\% , \tag{21}
$$

and

$$
D_{\mathrm{GA}}^{\mathrm{P}}=\frac{\left|I_{\mathrm{P}}-I_{\mathrm{GA}}\right|}{I_{\mathrm{P}}}\times 100\% , \tag{22}
$$

respectively.

## V. Simulations and Discussions

In this section, the spectral efficiencies of the downlink OFDMA cellular system are analyzed by using Monte Carlo simulations. We consider a synchronous, hexagonal OFDMA cellular network consisting of $7$ BSs as illustrated in Fig. 1. The cell radius is denoted by $R$. The MS moves along the solid line in Fig. 1. The corresponding path loss exponent is equal to 4, i.e., $\alpha_m = \alpha = 4$, $\forall m$. The transmit power of the BSs is assumed to be equal, i.e., $P_m = P$, $\forall m$, and it is assumed that the signal-to-noise ratio (SNR) is at 30 dB when the MS is at the cell-edge, i.e., $\gamma_m = \frac{PR^{-\alpha}}{\sigma_N^2} = 30\,\text{dB}$. The transmit signal on each subcarrier is real-valued and Gaussian-like, and the fading coefficients on each subcarrier are also real-valued and Gaussian-like. Finally, we assume the loading rates are identical in the system, i.e., $p_m = p$, $\forall m$.

Figure 6 shows the spectral efficiency, $I_{\text{P}}$, when the MS receiver has no CSI of the interference channels for $p = 0.5$. For comparison, we also plot the other two kinds of spectral efficiency, which have been introduced in the above section, i.e., $I_{\text{CSI}}$ and $I_{\text{GA}}$. It is shown that the three kinds of spectral efficiency are different with each others. $I_{\text{CSI}}$ is the highest one among the three spectral efficiencies, as it has full CSI of interference channels. $I_{\text{GA}}$ is the lowest one, because that the Gaussian-distributed interference is the worst case, in other words, the Gaussian distribution has maximum entropy, and thus leading to the severest signal distortion. It is also shown that when the MS moves toward the cell edge, the gap between $I_{\text{P}}$ and $I_{\text{GA}}$ increases. It is because when the MS receiver is at the center of the cell, the CCIs from adjacent BSs almost have equal power, thus the distribution of received total CCI is close to the Gaussian distribution. However, when the MS receiver is at the cell edge, the CCIs' powers are different, only few CCIs with higher power dominate the distribution. As we can see from Fig. 4, with the number of CCIs decreases, the deviation from Gaussian distribution increases. Therefore at the cell edge, the gap between $I_{\mathrm{p}}$ and $I_{\mathrm{GA}}$ is larger.

Figure 7 shows the difference factors, $D_{\mathrm{GA}}^{\mathrm{P}}$ and $D_{\mathrm{CSI}}^{\mathrm{P}}$, for different loading rates. It is shown that $D_{\mathrm{GA}}^{\mathrm{P}}$ increases significantly when the loading rate is small, or the receiver is near to the edge of the cell. It is because that, from Fig. 5, the distribution of total CCI signals significantly deviates from the Gaussian distribution as the loading rate decreasing or the distance between the BS and receiver increasing. For $D_{\mathrm{CSI}}^{\mathrm{P}}$, similar trend can also be found in Fig. 7. Another simulation result for LTE Urban macro-cell (UMa) scenario [23] is presented in Fig. 8. The system parameters and channel model are listed in Table I. It is shown that the trends of $D_{\mathrm{GA}}^{\mathrm{P}}$ and $D_{\mathrm{CSI}}^{\mathrm{P}}$ in Fig. 8 are similar to that shown in Fig. 7, but still have some slight differences from each other. The slopes of the curves in Fig. 7 are steeper than that in Fig. 8, especially when the receiver is near to the edge of the cell. The explanation of this result is that the channel models in the two simulations are different, more specifically, the different channel fading model leads to the different receive power of each interference signal from the adjacent BSs, and thus leads to the distribution of the summation of the interference signals, i.e., the distribution of CCI is also different from each other.

## VI. CONCLUSIONS

In this paper, the statistical properties of CCI and the spectral efficiency have been studied and analyzed for the downlink OFDM cellular system. The PDF of CCI signals has been derived with the considerations of path loss and multipath fading, which have been found to highly deviate from the Gaussian distribution. Based on the derived statistical model of CCI, the mutual information between the BS and the MS receiver has been derived. Simulation results have shown that the conventional spectral efficiency analysis based on the Gaussian-distributed interference model underestimates the potential capability of the OFDM cellular system, especially when the loading rate is small, or the receiver is near to the edge of the cell.

## APPENDIX A

### DERIVATION OF EQ. (2) AND EQ. (3)

Before deriving the PDF of CCI signal, we first introduce the following lemma.

***Lemma 1:*** Let $X_1 \sim N\left(0,\sigma_1^2\right)$ and $X_2 \sim N\left(0,\sigma_2^2\right)$ be independent Gaussian RVs. Then the product of $X_1$ and $X_2$, $X = X_1X_2$, has the following statistical properties with variance of $\sigma^2 = \sigma_1^2\sigma_2^2$ [21, Eq. (6.2) and Eq. (6.4)]

$$f_X\left(x\right) = \frac{1}{\pi\sigma}\mathrm{K}_0\left(\frac{|x|}{\sigma}\right), \tag{23}$$

and

$$\Psi_X\left(w\right) = \left(\frac{1}{1+\sigma^2w^2}\right)^{1/2}. \tag{24}$$

Note that $H_m$ and $X_m$ are two independent Gaussian RVs with unity variance, thus the PDF of $I_m = \Phi_m\sqrt{E_m}H_mX_m$ conditioned on $\Phi_m = 1$ and $E_m$ with variance of $\Phi_mE_m$ is given as $f_{I_m|\Phi_m}\left(x|\phi_m = 1\right) = \frac{1}{\pi\sqrt{E_m}}K_0\left(\frac{|x|}{\sqrt{E_m}}\right)$. For $\Phi_m = 0$, it can be seen that $I_m = 0$ and the PDF of $I_m$ is $f_{I_m|\Phi_m}\left(x|\phi_m = 0\right) = \delta\left(x\right)$. Therefore, the PDF of $I_m$ can be expressed as

$$\begin{aligned} f_{I_m}\left(x\right) &= \Pr\left(\phi_m = 1\right) f_{I_m|\Phi_m}\left(x|\phi_m = 1\right) + \Pr\left(\phi_m = 0\right) f_{I_m|\Phi_m}\left(x|\phi_m = 0\right) \\ &= p_m f_{I_m|\Phi_m}\left(x|\phi_m = 1\right) + \left(1-p_m\right) f_{I_m|\Phi_m}\left(x|\phi_m = 0\right) \\ &= \frac{p_m}{\pi\sqrt{E_m}}K_0\left(\frac{|x|}{\sqrt{E_m}}\right) + \left(1-p_m\right)\delta\left(x\right) \end{aligned}. \tag{25}$$

With the consideration of that the CF of $\delta\left(x\right)$ is $\Psi_\delta\left(w\right) = 1$ and Lemma 1, thus the CF of $I_m$ is given as

$$\Psi_{I_m}(w) = p_m \left( \frac{1}{1+E_m w^2} \right)^{1/2} + (1-p_m). \tag{26}$$

## APPENDIX B

### DERIVATION OF EQ. (11), EQ. (12), EQ. (13) AND EQ. (14)

Substituting $p_m = 1$ into Eq. (26) in Appendix A, the CF of one CCI signal can be expressed as

$$\Psi_{I_m}(w) = \left( \frac{1}{1+E_m w^2} \right)^{1/2}. \tag{27}$$

In the case of $M$ = 1, 4 and 6 CCI signals with the same average received power, the CF are given as $\frac{1}{1+E_m w^2}$, $\left(\frac{1}{1+E_m w^2}\right)^2$ and $\left(\frac{1}{1+E_m w^2}\right)^3$, respectively.

In the following analysis, we consider the generalized CF, i.e., $\left(\frac{1}{1+E_m w^2}\right)^n$ and $n$ is an integer. Using the inverse CF transform, the corresponding PDF can be expressed as

$$\begin{aligned} f_I(x) &= \frac{1}{2\pi}\int_{-\infty}^{\infty} \frac{1}{\left(1+E_1 w^2\right)^n} e^{-iwx} \mathrm{d}w \\ &= \frac{1}{2\pi}\int_{-\infty}^{\infty} \frac{1}{\left(1+E_1 w^2\right)^n} \left[\cos(-wx) + i\sin(-wx)\right] \mathrm{d}w. \\ &= \frac{1}{\pi}\int_{0}^{\infty} \frac{1}{E_1^n \left(\frac{1}{E_1} + w^2\right)^n} \cos(wx) \mathrm{d}w \end{aligned} \tag{28}$$

With the help of [22, Eq. (3.737)], Eq. (28) can be rewritten as

$$f_I(x) = \frac{\exp\left(-\frac{|x|}{\sqrt{E_1}}\right)}{2^{2n-1}(n-1)!\sqrt{E_1}} \sum_{k=0}^{n-1} \frac{(2n-k-2)!\left(2|x|\sqrt{\frac{1}{E_1}}\right)^k}{k!(n-k-1)!}. \tag{29}$$

Substituting $n = 1, 2$ and $3$, Eq. (29) reduces to Eq. (11), Eq. (12) and Eq. (13), respectively.

For $M$ = 6, and with the constraint of $E_m = E_{7-m}$ and $m = 1,2,3$, the PDF of $I = \sum_{m=1}^{6} I_m$ is given as

$$\begin{aligned}\Psi_I(w) &= \frac{1}{\left(1+E_1w^2\right)\left(1+E_2w^2\right)\left(1+E_3w^2\right)} \\ &= \frac{a_1}{1+E_1w^2} + \frac{a_2}{1+E_2w^2} + \frac{a_3}{1+E_3w^2}\end{aligned}, \tag{30}$$

where, $a_1 = \frac{E_1^2}{\left(E_1-E_2\right)\left(E_1-E_3\right)}$, $a_2 = \frac{E_2^2}{\left(E_2-E_1\right)\left(E_2-E_3\right)}$ and $a_3 = \frac{E_3^2}{\left(E_3-E_1\right)\left(E_3-E_2\right)}$. With consideration of Eq. (29) and Eq. (30), we finally derive Eq. (14).

TABLE:

TABLE I
PARAMETER OF LTE URBAN MACRO SCENARIO

| Parameter | Value |
|---|---|
| Deployment scenario | Urban macro-cell |
| Channel model | Urban macro model (LoS) |
| Inter-site distance | 500 m |
| Number of BSs | 19 |
| Carrier frequency | 2 GHz |
| System bandwidth | 10 MHz |
| Total BS transmit power | 46 dBm |
| Inter-site distance | 500 m |
| Thermal noise level | -174 dBm/Hz |

FIGURES:

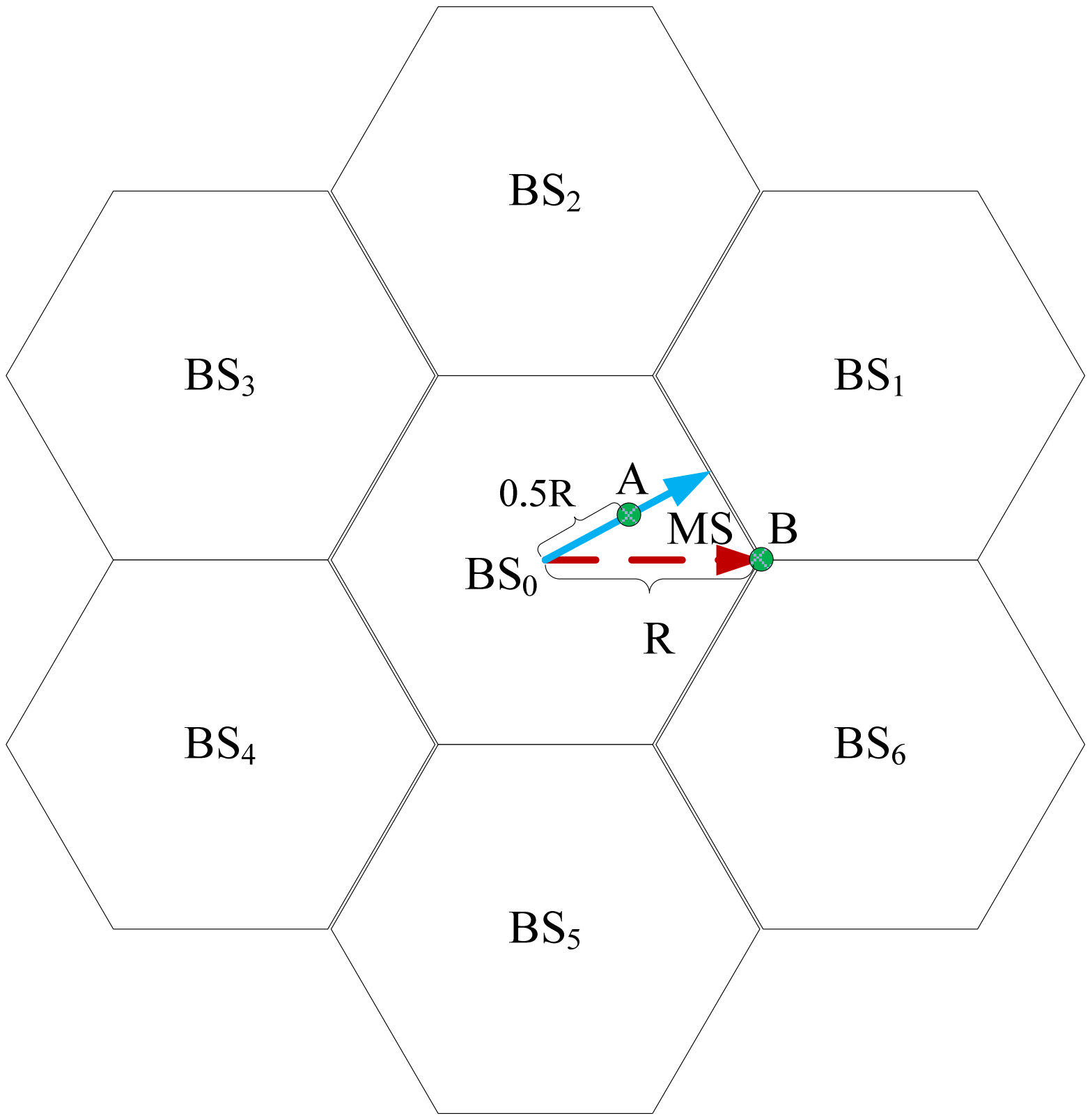

Fig. 1. Downlink, hexagonal, OFDMA cellular network for $M = 6$.

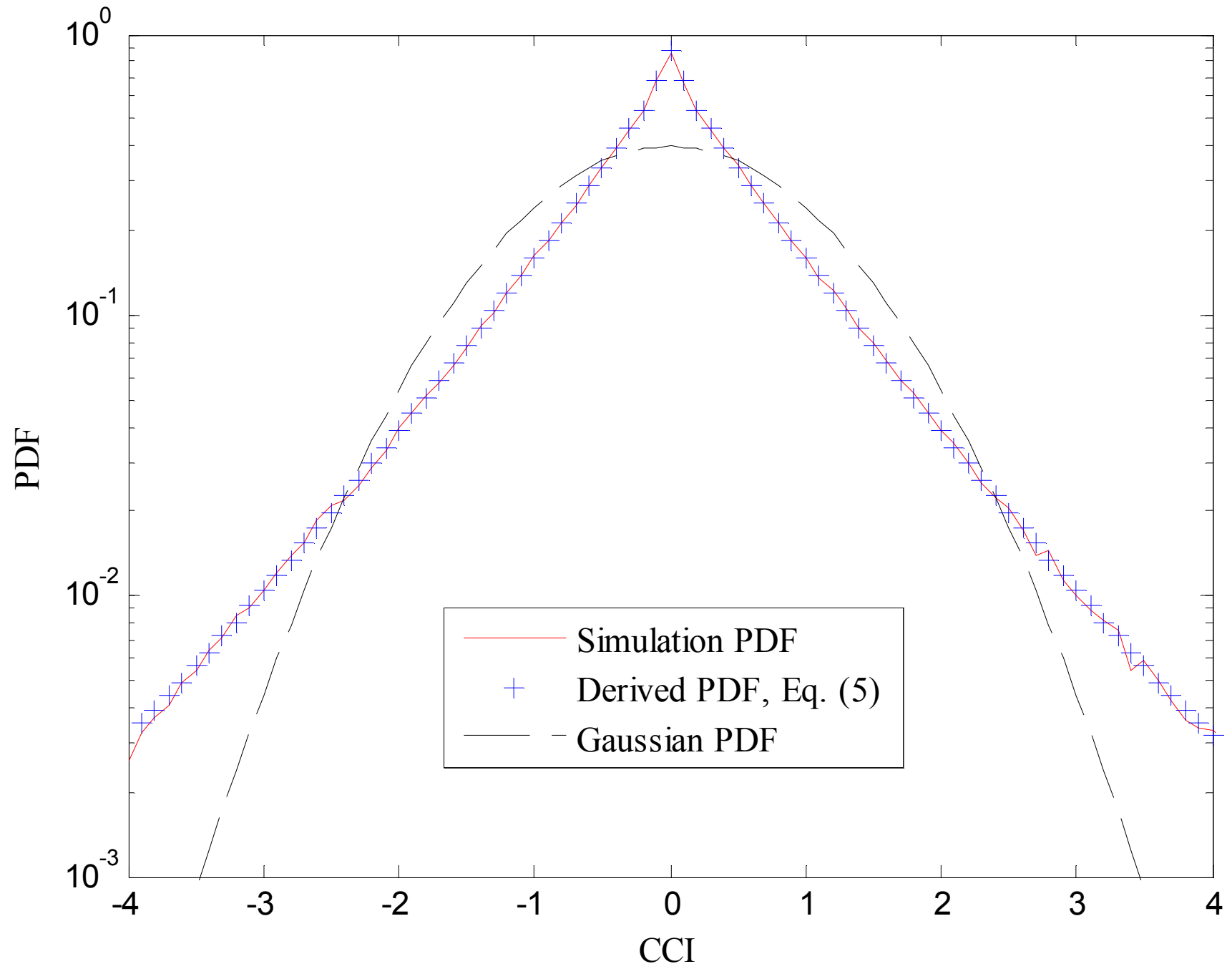

Fig. 2. PDF of CCI in downlink cellular OFDMA systems for $d = 0.5R$ and $p = 0.5$.

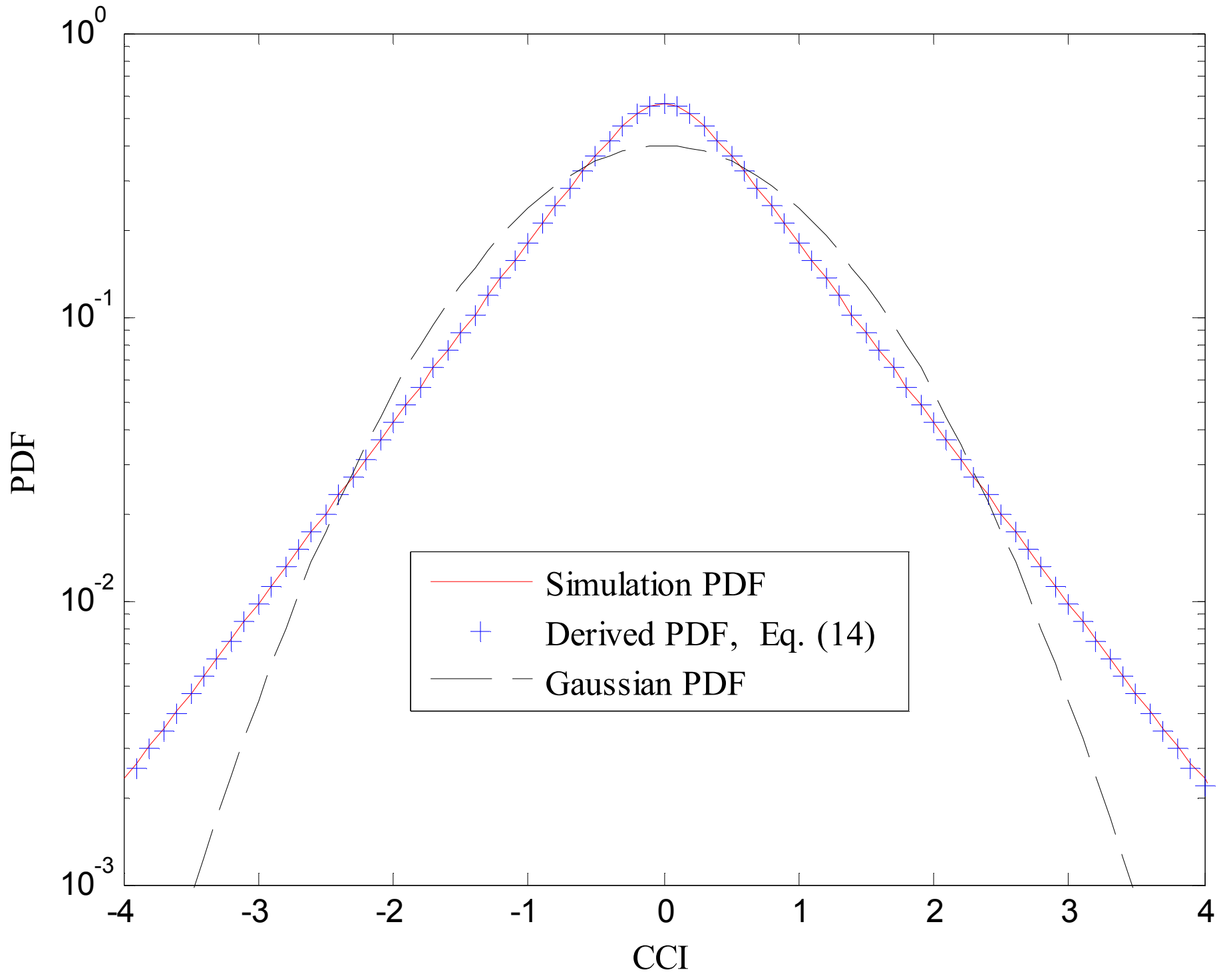

Fig. 3. PDFs of CCI in downlink cellular OFDMA systems for $d = R$ and $p = 1$.

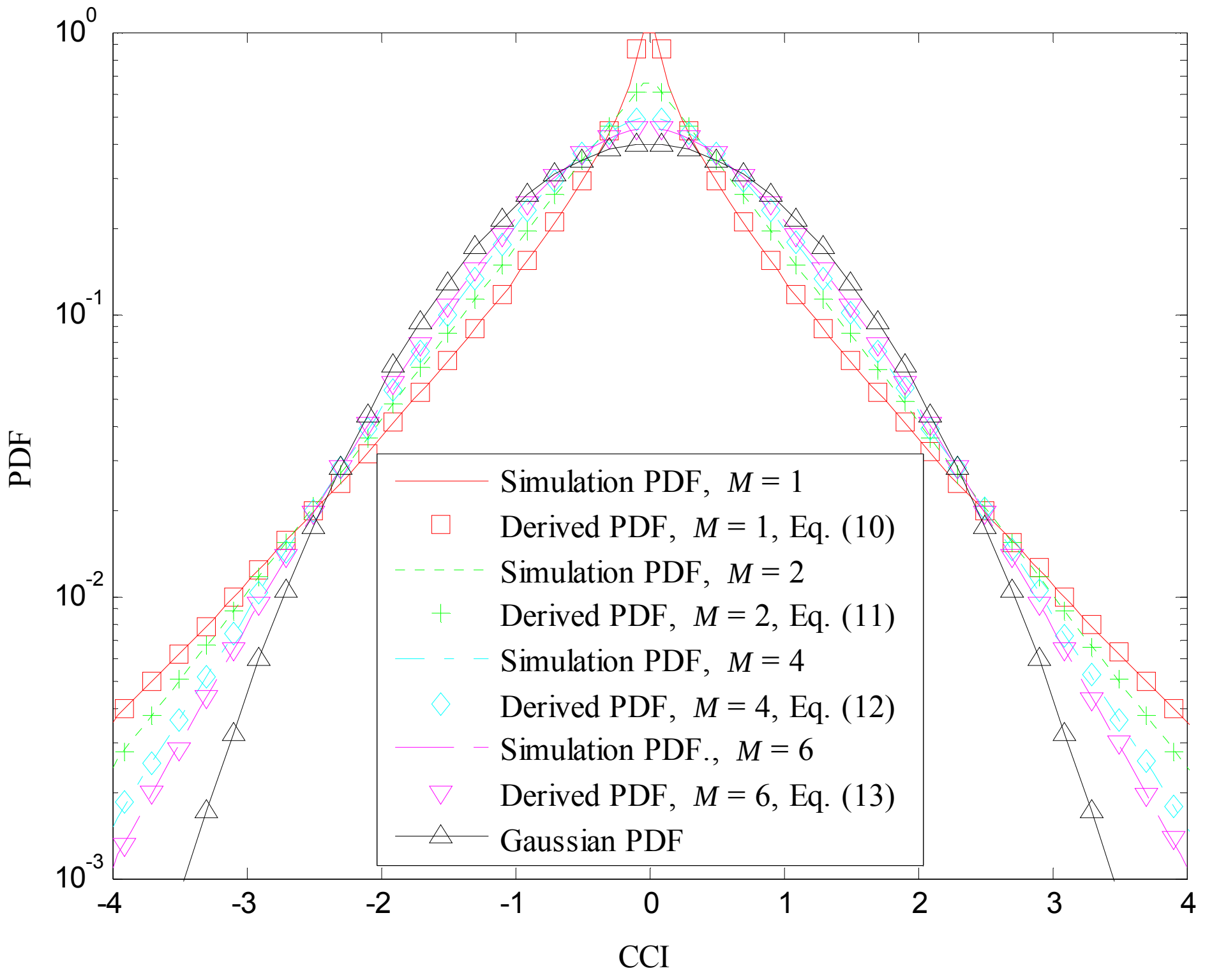

Fig. 4. PDFs of CCI for $M$ = 1, 2, 4 and 6.

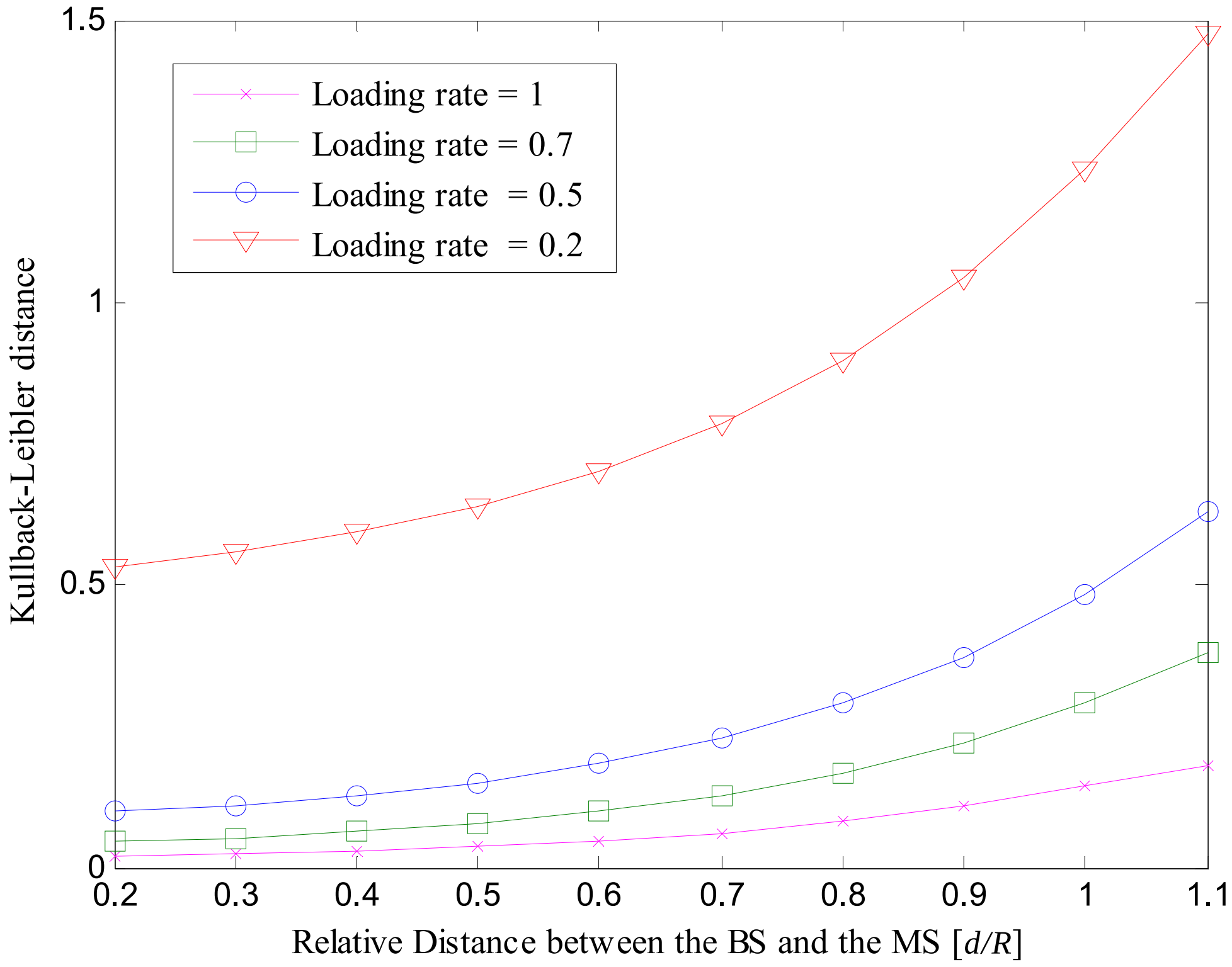

Fig. 5. Kullback-Leibler distance of the PDF of CCI and Gaussian distribution.

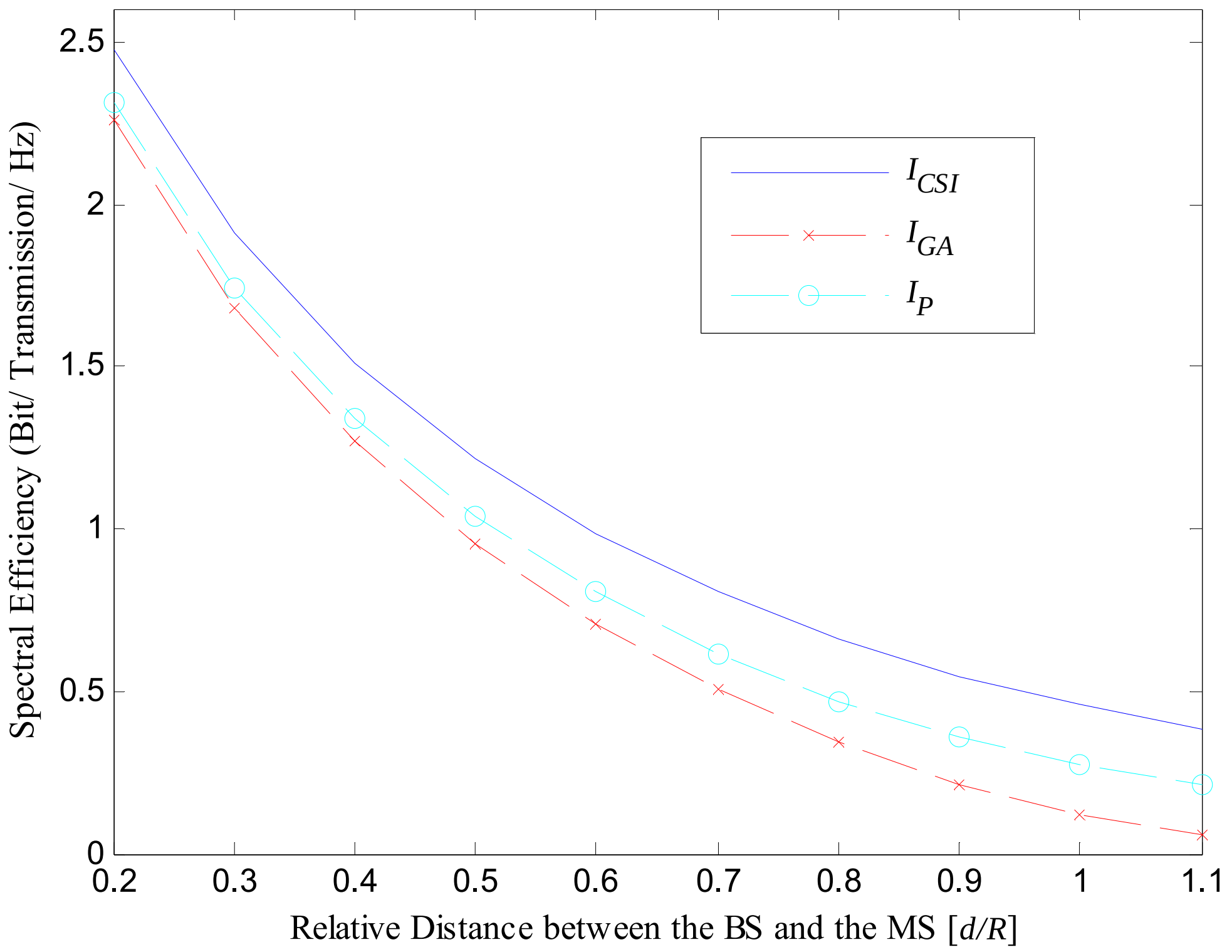

Fig. 6. Comparison of three kinds of spectral efficiency for $p = 0.5$.

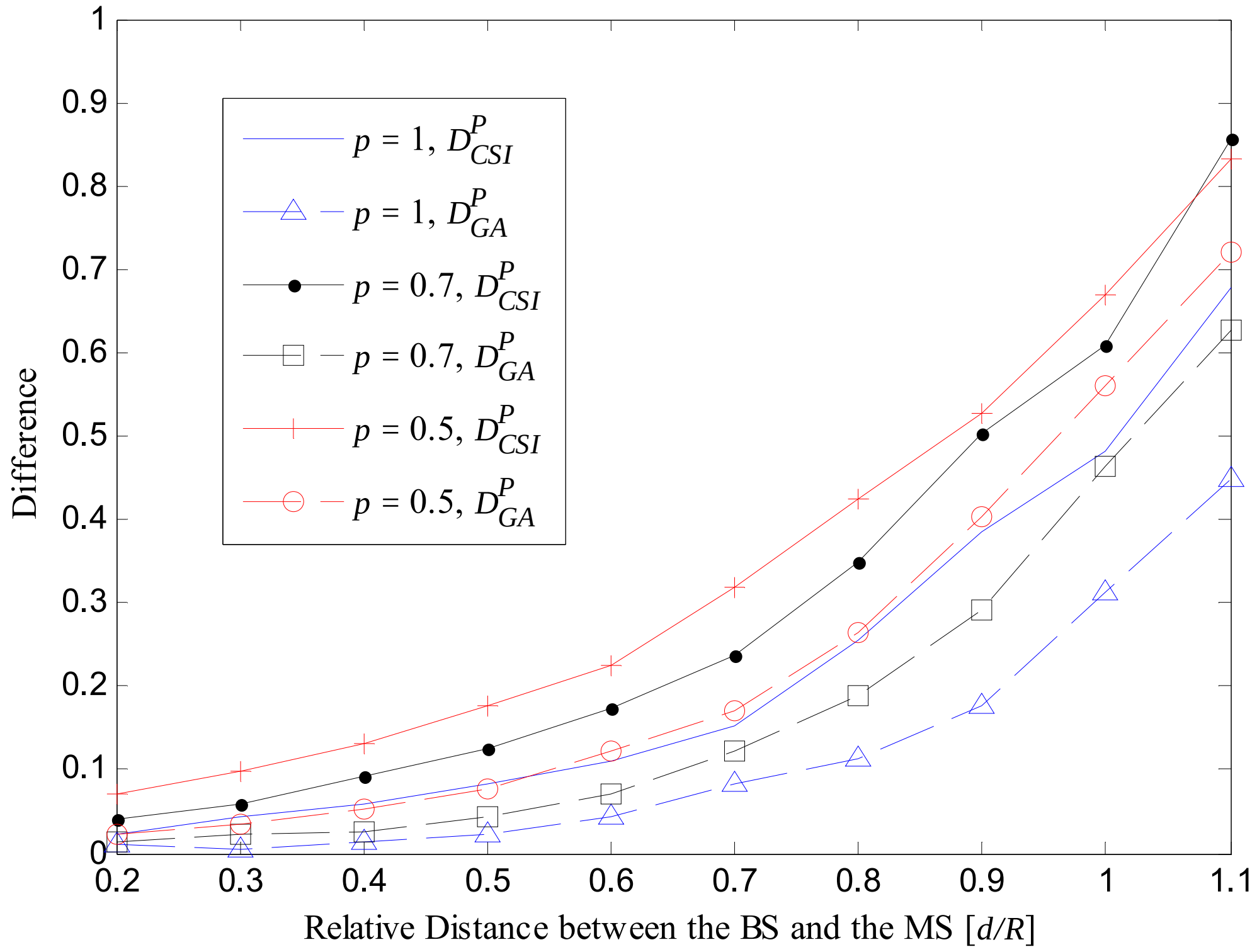

Fig. 7. Comparison of spectral efficiencies with different loading rates.

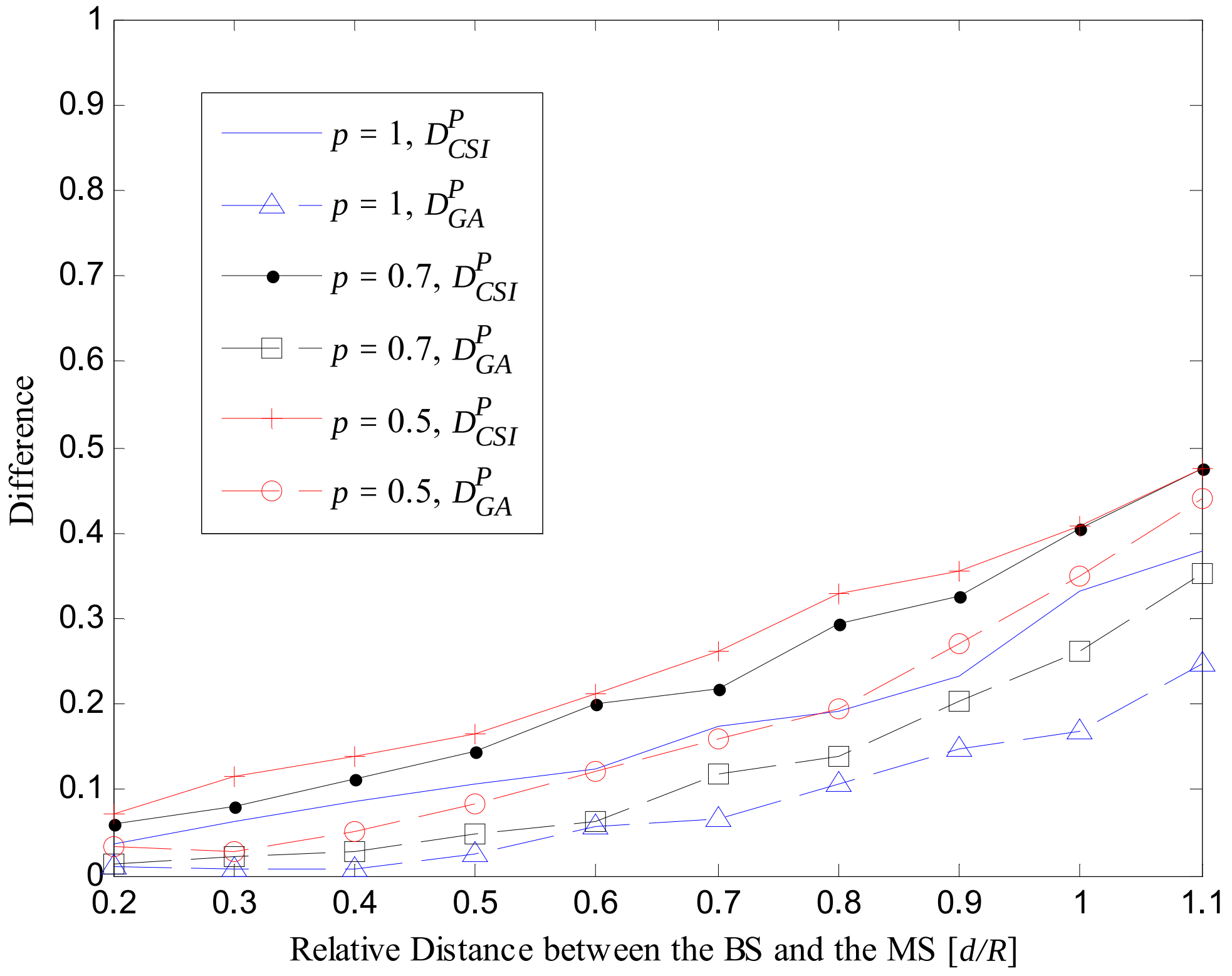

Fig. 8. Comparison of spectral efficiencies with different loading rates in LTE UMa scenario.